\documentclass[aps,pre,twocolumn,showpacs,superscriptaddress]{revtex4}
\usepackage{graphicx}

\begin{document}

\title{Network growth for enhanced natural selection}

\author{Valmir C. Barbosa}
\affiliation{Programa de Engenharia de Sistemas e Computa\c c\~ao, COPPE,
Universidade Federal do Rio de Janeiro,
Caixa Postal 68511, 21941-972 Rio de Janeiro - RJ, Brazil}

\author{Raul Donangelo}
\affiliation{Instituto de F\'\i sica,
Universidade Federal do Rio de Janeiro,
Caixa Postal 68528, 21941-972 Rio de Janeiro - RJ, Brazil}
\affiliation{Instituto de F\'\i sica, Facultad de Ingenier\'\i a,
Universidad de la Rep\'ublica,
Julio Herrera y Reissig 565, 11.300 Montevideo, Uruguay}

\author{Sergio R. Souza}
\affiliation{Instituto de F\'\i sica,
Universidade Federal do Rio de Janeiro,
Caixa Postal 68528, 21941-972 Rio de Janeiro - RJ, Brazil}
\affiliation{Instituto de F\'\i sica,
Universidade Federal do Rio Grande do Sul,
Caixa Postal 15051, 91501-970 Porto Alegre - RS, Brazil}

\begin{abstract}
Natural selection and random drift are competing phenomena for explaining the
evolution of populations. Combining a highly fit mutant with a population
structure that improves the odds that the mutant spreads through the whole
population tips the balance in favor of natural selection. The probability that
the spread occurs, known as the fixation probability, depends heavily on how the
population is structured. Certain topologies, albeit highly artificially
contrived, have been shown to exist that favor fixation. We introduce a
randomized mechanism for network growth that is loosely inspired in some of
these topologies' key properties and demonstrate, through simulations, that it
is capable of giving rise to structured populations for which the fixation
probability significantly surpasses that of an unstructured population. This
discovery provides important support to the notion that natural selection can be
enhanced over random drift in naturally occurring population structures.
\end{abstract}


\maketitle

Networks of agents that interact with one another underlie several important
phenomena, including the spread of epidemics through populations \cite{bbpv04},
the emergence of cooperation in biological and social systems
\cite{sp05,ohln06,tdw07}, the dynamics of evolution \cite{m58,lhn05}, and
various others \cite{gh05,sf07}. Typically, the dynamics of such interactions
involves the propagation of information through the network as the agents
contend to spread their influence and alter the states of other agents. In this
letter, we focus on the dynamics of evolving populations, particularly on how
network structure relates to the ability of a mutation to take over the entire
network by spreading from its node of origin.

In evolutionary dynamics, the probability that a mutation occurring at one of a
population's individuals eventually spreads through the entire population is
known as the mutation's fixation probability, $\rho$. In an otherwise
homogeneous population, the value of $\rho$ depends on the ratio $r$ of the
mutant's fitness to that of the other individuals, and it is the interplay
between $\rho$ and $r$ that determines the effectiveness of natural selection on
the evolution of the population, given its size. In essence, highly correlated
$\rho$ and $r$ lead to a prominent role of natural selection in driving
evolution; random drift takes primacy, otherwise \cite{n06}.

Let $P$ be a population of $n$ individuals and, for individual $i$, let $P_i$ be
any nonempty subset of $P$ that excludes $i$. We consider the evolution of $P$
according to a sequence of steps, each of which first selects $i\in P$ randomly
in proportion to $i$'s fitness, then selects $j\in P_i$ randomly in proportion
to some weighting function on $P_i$, and finally replaces $j$ by an offspring of
$i$ having the same fitness as $i$.

When $P$ is a homogeneous population of fitness $1$ (except for a randomly
chosen mutant, whose fitness is initially set to $r\neq 1$),
$P_i=P\setminus\{i\}$ \footnote{$\setminus$ denotes set difference.}, and
moreover the weighting function on every $P_i$ is a constant (thus choosing
$j\in P_i$ occurs uniformly at random), this sequence of steps is known as the
Moran process \cite{m58}. In this setting, evolution can be modeled by a simple
discrete-time Markov chain, of states $0,1,\ldots,n$, in which state $s$
indicates the existence of $s$ individuals of fitness $r$, the others $n-s$
having fitness $1$.

In this chain, states $0$ and $n$ are absorbing and all others are transient. If
$s$ is a transient state, then it is possible either to move from $s$ to $s+1$
or $s-1$, with probabilities $p$ and $q$, respectively, such that $p/q=r$, or
to remain at state $s$ with probability $1-p-q$. When $r>1$ (an advantageous
mutation), the evolution of the system has a forward bias; when $r<1$ (a
disadvantageous mutation), there is a backward bias. And given that the initial
state is $1$, the probability that the system eventually reaches state $n$ is
precisely the fixation probability, in this case denoted by $\rho_1$ and given
by
\begin{equation}
\rho_1=\frac{1-1/r}{1-1/r^n}
\end{equation}
(cf.\ \cite{n06}). The probability that the mutation eventually becomes extinct
(i.e., that the system eventually reaches state $0$) is $1-\rho_1$. Because
$\rho_1<1$, extinction is a possibility even for advantageous mutations.
Similarly, it is possible for disadvantageous mutations to spread through the
entirety of $P$.

In order to consider more complex possibilities for $P_i$, we introduce the
directed graph $D$ of node set $P$ and edge set containing every ordered pair
$(i,j)$ such that $j\in P_i$. The case of a completely connected $D$ (in which
every node connects out to every other node) corresponds to the Moran process.
But in the general case, even though it continues to make sense to set up a
discrete-time Markov chain with $0$ and $n$ the only absorbing states, analysis
becomes infeasible nearly always and $\rho$ must be calculated by computer
simulation of the evolutionary steps.

The founding work on this graph-theoretic perspective for the study of $\rho$
is \cite{lhn05}, where it is shown that we continue to have $\rho=\rho_1$ for
a much wider class of graphs. Specifically, the necessary and sufficient
condition for $\rho=\rho_1$ to hold is that the weighting function be such that,
for all nodes, the probabilities that result from the incoming weights sum up to
$1$ (note that this already holds for the outgoing probabilities, thus
characterizing a doubly stochastic process for out-neighbor selection). In
particular, if the weighting function is a constant for all nodes and a node's
in-degree (number of in-neighbors) and out-degree (the cardinality of $P_i$ for
node $i$, its number of out-neighbors) are equal to each other and the same for
all nodes, as in the Moran case, then $\rho=\rho_1$.

Other interesting structures, such as scale-free graphs \cite{ba99}, are also
handled in \cite{lhn05}, but the following two observations are especially
important to the present study. The first one is that, if $D$ is not strongly
connected (i.e., not all nodes are reachable from all others through directed
paths), then $\rho>0$ if and only if all nodes are reachable from exactly one of
$D$'s strongly connected components. Furthermore, when this is the case random
drift may be a more important player than natural selection, since fixation
depends crucially on whether the mutation arises in that one strongly connected
component. If $D$ is strongly connected, then $\rho>0$ necessarily.

The second important observation is that there do exist structures that
suppress random drift in favor of natural selection. One of them is the $D$ that
in \cite{lhn05} is called a $K$-funnel for $K\ge 2$ an integer. If $n$ is
sufficiently large, the value of $\rho$ for the $K$-funnel, denoted by $\rho_K$,
is
\begin{equation}
\rho_K=\frac{1-1/r^K}{1-1/r^{Kn}}.
\end{equation}
Thus, the $K$-funnel can be regarded as functionally equivalent to the Moran
graph with $r^K$ substituting for the fitness $r$. Therefore, the fixation
probability can be arbitrarily amplified by choosing $K$ appropriately, provided
$r>1$.

Noteworthy additions to the study of \cite{lhn05} can be found in
\cite{ars06,sar08}. In these works, analytical characterizations are obtained
for the fixation probability on undirected scale-free graphs, both under the
dynamics we have described (in which $j$ inherits $i$'s fitness) and the
converse dynamics (in which it is $i$ that inherits $j$'s fitness). The main
find is that the fixation probability is, respectively for each dynamics,
inversely or directly proportional to the degree of the node where the
advantageous mutation appears.

In this letter, we depart from all previous studies of the fixation probability
by considering the question of whether a mechanism exists for $D$ to be grown
from some simple initial structure in such a way that, upon reaching a
sufficiently large size, a value of $\rho$ can be attained that substantially
surpasses the Moran value $\rho_1$ for an advantageous mutation. Such a $D$
might lack the sharp amplifying behavior of structures like the $K$-funnel, but
being less artificial might also relate more closely to naturally occurring
processes. We respond affirmatively to the question, inspired by the observation
discussed above on the strong connectedness of $D$, and using the $K$-funnel as
a sieving mechanism to help in looking for promising structures. It should be
noted, however, that since other amplifiers exist with capabilities similar to
those of the $K$-funnel (e.g., the $K$-superstar \cite{lhn05}), alternatives to
the strategy we introduce that are based on them may also be possible.

In a $K$-funnel, nodes are organized into $K$ layers, of which layer $k$
contains $b^k$ nodes for some fixed integer $b\ge 2$ and $k=0,1,\ldots,K-1$. It
follows that the $K$-funnel has $(b^K-1)/(b-1)$ nodes. A node in layer $k$
connects out to all nodes in layer $k-1$ (modulo $K$, so that an edge exists
directed from the single node in layer $0$ to each of the $b^{K-1}$ nodes in
layer $K-1$). A $K$-funnel is then, by construction, strongly connected. For a
given value of $n$, our strategy for growing $D$ is to make it a layered graph
like the $K$-funnel, but letting it generalize on the $K$-funnel by allowing
each layer to have any size (number of nodes), provided no layer remains empty.

Graph $D$ is the graph that has $n$ nodes in the the sequence $D_0, D_1,\ldots$
of directed graphs described next. Graph $D_0$ has $K$ layers, numbered $0$
through $K-1$, each containing one node. The node in layer $k$ connects out to
the node in layer $k-1$ (modulo $K$). For $t\ge 0$ an integer, $D_{t+1}$ is
obtained from $D_t$ by adding one new node, call it $i$, to a randomly chosen
layer, say layer $k$, according to a criterion to be discussed shortly. Node $i$
is then connected out to all nodes in layer $k-1$ (modulo $K$) and all nodes in
layer $k+1$ (modulo $K$) are connected out to node $i$. Graph $D_t$ is then
strongly connected for all $t$. We note that there are as many possibilities for
the resulting $D$ as for partitioning $n$ indistinguishable objects into $K$
nonempty, distinguishable sets arranged circularly, provided we discount for
equivalences under rotations of the sets. A lower bound on this number of
possibilities is ${n\choose K}/n$, which for $K\ll n$ is roughly $n^{K-1}/K!$.

Before we describe the rule we use to decide which layer is to receive the new
node, $i$, it is important to realize that the double stochasticity mentioned
earlier implies that $\rho=\rho_1$ for $D_0$. However, this ceases to hold
already for $D_1$ and may not happen again as the graph gets expanded. So,
whatever the rule is, we are aiming at higher $\rho$ values by giving up on the
doubly stochastic character of the process whereby fitness propagates through
the graph.

For $t\ge 0$ and $k$ any layer of $D_t$, if we consider the layers in the
upstream direction from $k$, we call $k^+$ the first layer we find whose
successor has at most as many nodes as itself. In particular, if the successor
of layer $k$ does not have more nodes than $k$, then $k^+=k$. Now let $d(k^+,k)$
be the distance from layer $k^+$ to layer $k$ in $D_t$ (i.e., the number of
edges on a shortest directed path from any node in $k^+$ to any node in $k$).

Layer $k$ is selected to receive node $i$ to yield $D_{t+1}$ with probability
\begin{equation}
p_k\propto[K-d(k^+,k)]^a
\end{equation}
for some $a\ge 1$. This criterion is loosely suggested by the topology of the
$K$-funnel. It seeks to privilege first the growth of each layer $\ell$ such
that $k^+=\ell$ for some $k$, then the growth of the layer $k$ that is
immediately downstream from $\ell$, provided $k^+=\ell$, and so on through the
other downstream layers.

In our simulations we use $n\le 1\,000$ nearly exclusively and grow a large
number of $D$ samples. The calculation of $\rho$ for a given $D$ involves
performing several independent simulations (we use $10\,000$ in all cases), each
one starting with the fitness-$r$ mutant substituting for any of the $n$ nodes
and proceeding as explained earlier until the mutation has either spread through
all of $D$'s nodes or died out (we use constant weighting throughout). The
fraction of simulations ending in fixation is taken as the value of $\rho$ for
that particular $D$. This calculation can be very time-consuming, so we have
adopted a mechanism to decide whether to proceed with the calculation for a
given $D$ or to discard it.

Our mechanism is based on establishing a correlation threshold beyond which $D$
is declared sufficiently similar to the $K$-funnel to merit further
investigation. The measure of correlation that we use is the Pearson correlation
coefficient between two sequences of the same size, which lies in the interval
$[-1,1]$ and indicates how closely the two sequences are to being linearly
correlated (a coefficient of $1$ means a direct linear dependence). For
sequences $X$ and $Y$, the coefficient, denoted by $C(X,Y)$, is given by
$C(X,Y)=\mathrm{cov}(X,Y)/\sigma_X\sigma_Y$, where $\mathrm{cov}(X,Y)$ is the
covariance of $X$ and $Y$, $\sigma_X$ and $\sigma_Y$ their respective standard
deviations.

In our case, $X$ and $Y$ are length-$K$ sequences. If we renumber the layers of
$D$ so that the layer with the greatest number of nodes becomes layer $K-1$, the
one immediately downstream from it layer $K-2$, and so on through layer $0$,
then we let the sequences $X$ and $Y$ be such that $X_k=k$ and $Y_k=\ln n_k$,
where $n_k$ is the number of nodes in layer $k$. Notice that, when $D$ is the
$K$-funnel itself, then $n_k=b^k$ with $b\ge 2$, whence $Y_k=(\ln b)X_k$ and
$C(X,Y)=1$.

Every $D$ whose sequences $X$ and $Y$ lead $C(X,Y)$ to surpass the correlation
threshold is as close to having $n_k$ given by some exponential of $k$ as the
threshold allows. However, the near-linear dependence of the two sequences is
not enough, since the base of such an exponential, which we wish to be as large
as possible, can in principle be very small (only slightly above $1$), for very
gently inclined straight lines. On the other hand, a steeper straight line
indicates a faster reduction of layer sizes as we progressively move toward
layer $0$ from layer $K-1$ through the other layers. In the analysis that
follows, then, we also use the slope of the least-squares linear approximation
of $Y$ as a function of $X$, denoted by $S(X,Y)$ and given by
$S(X,Y)=\mathrm{cov}(X,Y)/\sigma_X^2$. For $C(X,Y)$ close to $1$, the base of
the aforementioned exponential approaches $e^{S(X,Y)}$.

\begin{figure}[t]
\vspace{0.35cm}
\includegraphics[scale=0.31]{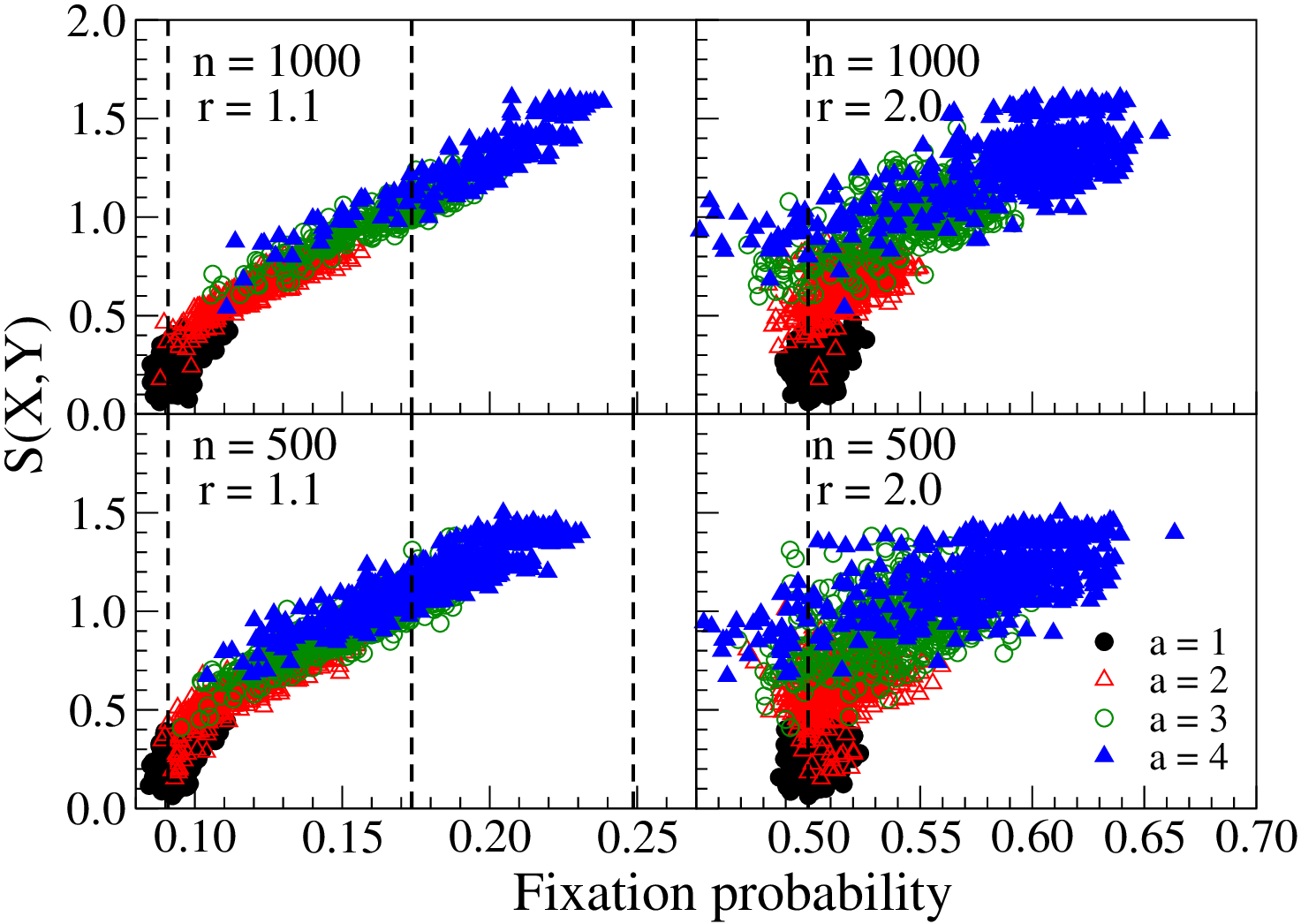}
\caption{(Color online) Simulation results for $K=5$. Each graph $D$ for which
$C(X,Y)>0.9$ is represented by its fixation probability and by the slope
$S(X,Y)$. For each combination of $n$ and $r$, $500$ graphs are shown,
corresponding roughly to $12\%$ of the number of graphs that were grown. Dashed
lines mark $\rho_1$ through $\rho_3$ for $r=1.1$, $\rho_1$ for $r=2.0$.}
\label{fig:5layers}
\end{figure}

Our simulation results are summarized in Fig.~\ref{fig:5layers}, where $K=5$,
$n=500,1\,000$, and $r=1.1,2.0$. For each combination and each of four $a$
values ($a=1,2,3,4$), a scatter plot is given representing each of the graphs
generated by its fixation probability and the slope $S(X,Y)$ for its two
sequences, provided $C(X,Y)>0.9$. We see that, in all cases, strengthening the
layer-selection criterion by increasing $a$ has the effect of moving most of the
resulting graphs away from the Moran probability ($\rho_1$) and also away from
the near-$0$ slope.

It is important to notice that, in the absence of the slope indicator for each
graph, we would be left with a possibly wide range of fixation probabilities for
the same value of $a$, unable to tell the true likeness of the best graphs to
the $K$-funnel without examining their structures one by one. In a similar vein,
the results shown in Fig.~\ref{fig:5layers} emphasize very strongly the role
of our particular choice of a rule for selecting layers, as opposed to merely
proceeding uniformly at random. To see this, it suffices that we realize that
uniformly random choices correspond to setting $a=0$ in the expression for
$p_k$, and then we can expect the graphs that pass the correlation threshold to
be clustered around the points of $\rho\sim \rho_1$ and $S(X,Y)\sim 0$.

\begin{figure}[t]
\vspace{0.35cm}
\includegraphics[scale=0.31]{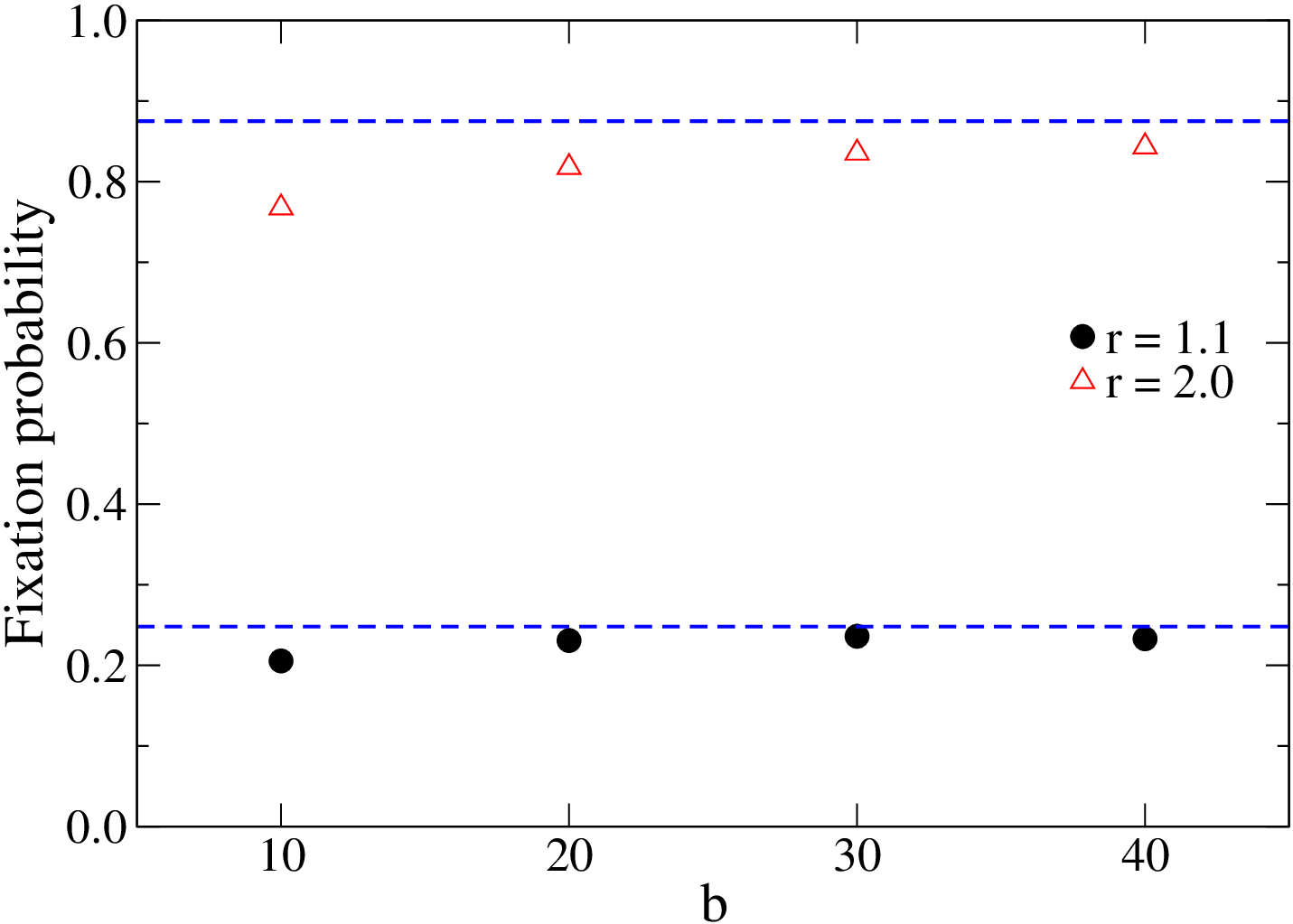}
\caption{(Color online) Simulation results for the $3$-funnel. Dashed lines mark
the values of $\rho_3$.}
\label{fig:funnel}
\end{figure}

We also note a sharp variation in how the fixation probabilities of the graphs
relate to the asymptotic fixation probabilities of the $K$-funnel as a mutant's
fitness is increased. For $r=1.1$, the graphs exhibiting the highest fixation
probabilities, and also the highest slopes, are such that $\rho$ is somewhere
between $\rho_2$ and $\rho_3$. For $r=2.0$, though, this happens between
$\rho_1$ and $\rho_2$ ($=0.75$, not shown), therefore providing considerably
less amplification. Part of the reason why this happens may be simply that the
more potent amplifiers are harder to generate by our layer-selection mechanism
as $r$ is increased. But it is also important to realize that, even for the
$K$-funnel, achieving a fixation probability near $\rho_K$ requires
progressively larger graphs as $r$ is increased. This is illustrated in
Fig.~\ref{fig:funnel} for $K=3$ and the same two values of $r$.

\begin{figure}[h]
\vspace{0.60cm}
\includegraphics[scale=0.31]{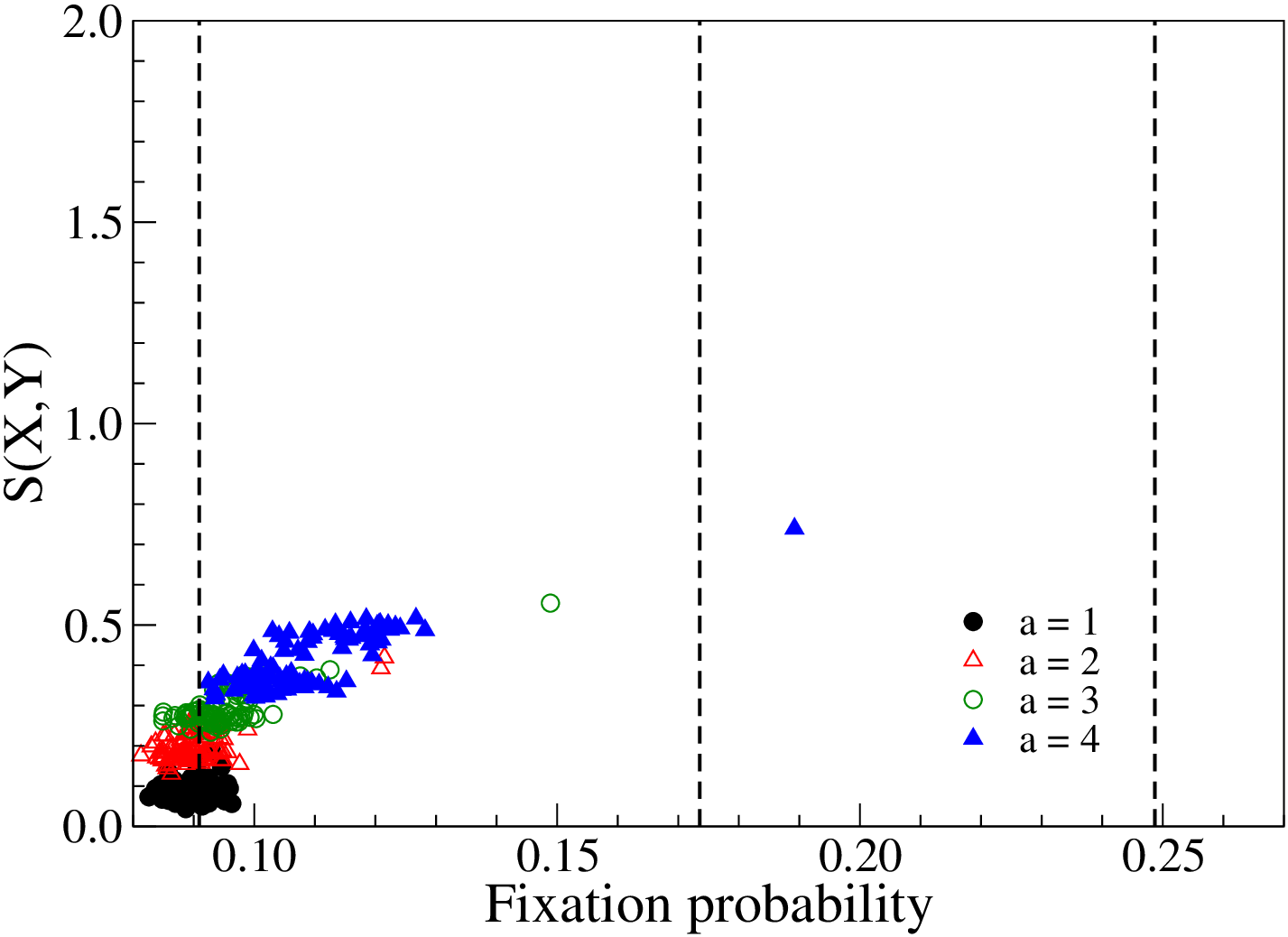}
\caption{(Color online) Simulation results for $K=10$, $n=10\,000$, and $r=1.1$.
Each graph $D$ having $C(X,Y)>0.9$ is represented by its fixation probability
and by the slope $S(X,Y)$. There are $100$ graphs, corresponding roughly to
$0.04\%$ of the graphs that were grown. Dashed lines mark $\rho_1$ through
$\rho_3$.}
\label{fig:10layers}
\end{figure}

Additional simulation results, for the much larger case of $K=10$ and
$n=10\,000$, are presented in Fig.~\ref{fig:10layers} for $r=1.1$ and
$a=1,2,3,4$. Computationally, this case is much more demanding than those of
Fig.~\ref{fig:5layers}, owing mainly to the number of distinct networks that
can occur, as discussed earlier (in fact, for $K=10$ and $n=10\,000$, this
number is at least of the order of $10^{33}$). Consequently, many fewer graphs
surpassing the $0.9$ correlation threshold were obtained. Even so, one possible
reading is that results similar to those reported in Fig.~\ref{fig:5layers}
can be expected, but this remains to be seen.

In summary, we have demonstrated that strongly connected layered networks can be
grown for which the fixation probability significantly surpasses that of the
Moran process. The growth mechanism we use aggregates one new node at a time and
chooses the layer to be enlarged by the addition of the new node as a function
of how far layers are from those whose populations are the closest upstream
local maxima. A great variety of networks can result from this process, but we
have shown that correlating each resulting $K$-layer network with the $K$-funnel
appropriately works as an effective filter to pinpoint those of distinguished
fixation probability. Further work will concentrate on exploring other growth
methods and on targeting the growth of more general structures.

We acknowledge partial support from CNPq, CAPES, FAPERJ BBP grants, and the
joint PRONEX initiative of CNPq and FAPERJ under contract 26.171.528.2006.

\bibliography{fixprob}
\bibliographystyle{apsrev}

\end{document}